\documentstyle[prd,aps]{revtex}

\def\bea{\begin{eqnarray}}
\def\eea{\end{eqnarray}}
\def\fr{\frac}
\def\l{\left}
\def\pa{\partial}

\def\r{\right}

\def\cs{c\!\!/ }
\def\ss{s\!\!\!/ }

\def\cl{{\cal L}}
\def\dag{\dagger}
\def\eps{\epsilon}
\def\veps{\varepsilon}
\def\om{\omega}
\def\nn{\nonumber}
\def\Tr{{\,\mbox{Tr}}}

\begin{document}
\draft

\preprint{TAN-FNT-99-03}

\title{Multibaryons with heavy flavors in the Skyrme model}
 
\author{Carlos L. Schat} 
\address{ Centro Brasileiro de Pesquisas F\'{\i}sicas , DCP,  \\
Rua Dr. Xavier Sigaud 150, 22290-180, Rio de Janeiro, RJ, Brazil. }

\author{Norberto N. Scoccola}
\address{ Physics Department, Comisi\'on Nacional de Energ\'{\i}a At\'omica,\\
	  Av.Libertador 8250, (1429) Buenos Aires, Argentina.\\
 Universidad Favaloro, Sol{\'{\i}}s 453, (1078) Buenos Aires, Argentina.}

\maketitle
\begin{abstract}
\noindent
We investigate the possible existence of multibaryons with heavy flavor
quantum numbers using the bound state approach to the topological soliton 
model and the recently proposed approximation for multiskyrmion fields based on 
rational maps. We use an effective interaction lagrangian which consistently incorporates both chiral 
symmetry and the heavy quark symmetry including the corrections up to order ${\cal O}(1/m_Q)$. 
The model predicts some narrow heavy flavored multibaryon states with baryon 
number four and seven.  

\end{abstract}


%Keywords : heavy flavor multiskyrmion , rational map , heavy quark symmetry  

\section{Introduction}

In recent years multibaryons have received a considerable amount of attention. A reason for 
this is that their hypothetical
existence could provide valuable information about the nature of the strong interactions
at low energies. Perhaps the most celebrated example is that of the H-dibaryon predicted by
Jaffe more than twenty years ago\cite{Jaf77}. Since then the possible existence of some 
other exotic states has been investigated in various models. Of particular
interest are those containing flavor (i.e. ${\cal S}$ , ${\cal C}$ , ${\cal B}$ ) quantum numbers. In fact, it has 
been speculated that strange matter could be stable\cite{Wit84}. This has lead to numerous 
investigations of the properties of strange matter in bulk and in finite lumps (for a 
recent review see Ref.\cite{GSB98}). Moreover, 
with the advent of heavy ion colliders there is now the possibility of producing 
strange\cite{E864} and even charmed\cite{SV98} multibaryonic states with rather low baryon 
number in the laboratory. These new developments provide further motivation for the study of 
the properties of multibaryons with heavy flavor quantum numbers. 
Most of the known predictions come from MIT bag model (see for example Refs.\cite{AMS78}) 
or non-relativistic quark
model based calculations. Here, we will adopt a different point of view. We will asume
that the heavy flavor multibaryons are formed by an $SU(2)$ multiskyrmion with some
heavy mesons bound to it. This is basically an extension of the bound state approach
to strange hyperons originally introduced by Callan and Klebanov\cite{CK85} and later shown
to describe heavier flavor baryons as well\cite{RRS90}. A study of strange multibaryons
within this approach has been recently presented in Ref.\cite{SS98}. There, a chiral lagrangian 
written in terms of pseudoscalar meson fields with some chiral symmetry breaking terms has 
been used as the effective lagrangian. As well known by now, although adequate for the light
(up and down) and strange sectors, such type of effective lagrangian has to be modified
when heavier flavors (e.g. charm) are incorporated. In that case, 
heavy quark symmetry\cite{IW88} has to be imposed. This symmetry requires that both
the heavy pseudoscalar and the heavy vector fields appear explicitly in the effective
lagrangian. Lagrangians which have both chiral symmetry and heavy quark symmetry have
been described in the literature\cite{BD92,Yan92}. In our calculation we will adopt such type of lagrangian. 
As already 
mentioned, in our description the baryon number of the system comes from a non-trivial
soliton configuration in the light sector. Until very recently only few multiskyrmion
configurations (i.e. those with $B \leq 4$) were known. In 1997, however,
after some demanding numerical work Battye and Sutcliff \cite{BS97} were able to identify those
which are believed to be the lowest energy configurations with baryon number up to $B=9$.
Interestingly, all these configurations have the symmetries corresponding to the
regular polyhedra. Even more important
for our purposes, Houghton, Manton and Sutcliffe \cite{HMS98}
have exploited the similarities between the BPS monopoles and skyrmions
to propose some ans\"atze based on rational maps.
They have shown that
such configurations approximate very well the numerically found lowest
energy solutions with $B \leq 9$. In our investigations we will make
use of these approximate ans\"atze. An interesting feature of these configurations
is that for $B > 1$ the derivative of the radial soliton profile vanishes
at the origin. Since to leading order in the inverse of the heavy quark mass $m_Q$ the heavy 
meson-soliton interaction is proportional to this quantity\cite{JMW93} we do not expect
any bound state in that approximation. However, next-to-leading order corrections in $1/m_Q$ 
are required even to describe the spectrum of $B=1$ heavy baryons\cite{OPM94}. In the present 
work these corrections will be properly taken into account. 
    
This article is organized as follows. In Sec.II we introduce the effective lagrangian
together with the ansatz for the multiskyrmion configurations. In Sec.III we
present our numerical results. Finally, in Sec.IV our main conclusions are given.

\section{The Lagrangian}

To describe the dynamics of the light and massive mesons interacting with 
each other we will consider an effective lagrangian 
of the following form \cite{Yan92}
\bea \label{hlag}
\cl  &=& \cl_{l} + (D_\mu \phi)^\dag D^\mu \phi - m^2_\phi \phi^\dag \phi 
- \fr{1}{2} 
\psi^{\dag}_{\mu\nu} \psi^{\mu\nu} + m^2_\psi \psi^{\dag}_\mu
\psi^{\mu} \ \nn \\
& & + i f \l( \phi^\dag a^\mu \psi _\mu + \psi_\mu^{\dag} a^\mu
\phi \r) 
+ \fr{i}{2} \ g \ \eps^{\mu\nu\rho\sigma} \l(\psi^{\dag}_{\mu\nu} a_\rho
\psi _\sigma
+ \psi_\sigma^{\dag} a_\rho \psi _{\mu\nu} \r) \ ,  
\label{fulllag}   
\eea
where $\cl_l$ is the effective light meson lagrangian. In the present
work we will consider pions as the only explicit light degree of freedom
and choose $\cl_l$ to be  simply the  Skyrme lagrangian.
Consequently,   the effective lagrangian  $\cl_l$ ,  written in terms of the chiral field 
$U = \exp( i \vec \tau \cdot \vec \pi /f_\pi )$,
reads
\begin{eqnarray}
{\cal L}_l &=& {f^2_\pi \over 4} \Tr\l[\partial_\mu U^\dagger \partial^{\mu} U \r] + 
{1\over{32 \eps^2}} \Tr \Big[ [U^\dagger\partial_\mu U, U^\dagger\partial_\nu U]^2 \Big] 
+ {f^2_\pi m_\pi^2\over4} \Tr\l[ U + U^\dagger - 2 \r] \ . 
\end{eqnarray}
Here, $f_\pi$ is the pion decay constant and $\eps$ is the so-called Skyrme 
parameter. In Eq.(\ref{fulllag}), $\phi$ represents the heavy pseudoscalar doublet
and 
$\psi_\mu$ the corresponding vector doublet. For example, for charmed mesons 
\begin{equation}
\phi = \left(\begin{array}{l} \bar D^0 \\ D^- \end{array}\right)
\qquad \quad ; \qquad \quad
\psi = \left(\begin{array}{l} \bar D^{*0} \\ D^{*-} \end{array}\right) \ .
\end{equation}
Moreover, $f$ and $g$ are the $\phi\psi_\mu\pi$ and
$\psi_\mu\psi_\mu\pi$ coupling constants, respectively, and
\begin{eqnarray}
D_\mu &=& \pa_\mu + v_\mu  \ , \\
\psi_{\mu\nu} &=& D_\mu \psi_\nu - D_\nu \psi_\mu \ .
\end{eqnarray}
Finally, in terms of the chiral field the currents $v_\mu$ and $a_\mu$ read
\begin{eqnarray}
v_\mu &=& \frac{1}{2} \left( \sqrt{U^\dagger} \partial_\mu \sqrt{U} + \sqrt{U} \partial_\mu \sqrt{U^\dagger}
                      \right) \ , \\
a_\mu &=& \frac{1}{2} \left( \sqrt{U^\dagger} \partial_\mu \sqrt{U} - \sqrt{U} \partial_\mu \sqrt{U^\dagger}
                      \right) \ .
\end{eqnarray}

As usual in the bound state model we should first determine the static skyrmion background. 
For this purpose we introduce the rational map ansatz for the pion field. It reads\cite{HMS98}
\begin{equation}
\vec \pi = f_\pi F(r) \ \hat n \ .
\label{PS}
\end{equation}
Here, $F(r)$ is the (multi)skyrmion profile which depends on the radial coordinate
only and $\hat n$ is a unit vector given by
\begin{equation}
\hat n = {1\over{1+|R|^2}} \left( 2 \ \Re(R) \ \hat \imath +
				      2 \ \Im(R) \ \hat \jmath +
				     ( 1 - |R|^2 ) \ \hat k \right) \ ,
\label{pians}
\end{equation}
where $R$ is the rational map corresponding to a certain winding number $B$ which is identified with the baryon number.
Such map is usually written as a function of the complex variable $z$
which is related to the usual spherical coordinates $\theta, \varphi$ via
stereographic projection, namely $z = \tan(\theta/2) \exp(i \varphi)$.
For example, the map corresponding to the $B=1$ hedgehog ansatz
is the identity map $R = z$. The explicit form of the maps corresponding
to the other baryon numbers $B \leq 9$ can be found in Ref.\cite{HMS98}. 
Using Eq.(\ref{PS}) it is possible to obtain the expressions of the
$a_\mu$ and $v_\mu$ currents to leading order in $N_c$. The time components vanish
at this order while the space components result 
\bea
a^i &=& -{i\over2} \left[ F' \ \vec \tau \cdot \hat n \ \hat r^i +
s \ \vec \tau \cdot \nabla^i \hat n   \right] \ , 
\\
v^i &=& - i \ss^2 \ \left( \hat n \times \nabla^i \hat n
\right)  
\cdot \vec \tau  \ , 
\eea
where we have introduced the short hand notation $s = \sin F$, $\ss = \sin (F/2)$. 
The radial profile function $F(r)$ is determined by minimizing the soliton
energy. Details of this procedure as well as plots of these 
functions for different baryon numbers are given in Ref.\cite{HMS98}.

To order $N_c^0$, we have a system of heavy mesons moving in the static soliton 
background. To derive the explicit form of the relevant heavy meson-soliton 
lagrangian we need some consistent ans\"atze for the heavy meson fields. For the 
pseudoscalar field we use\cite{SS98} 
\bea 
\phi(\vec r, t) &=& \fr{1}{\sqrt{4 \pi}} \ \phi(r,t) \ 
\vec \tau \cdot \hat n \ \chi  \ , 
\label{HPS}
\eea
where $\chi$ is a two-component spinor. To obtain the corresponding ansatz for the heavy vector meson 
field it is convenient to
analize the coupling terms in the effective lagrangian. They are the last two terms 
in Eq.(\ref{fulllag}). From their structure and the form of the $a_\mu$ and $v_\mu$ currents
in the static limit it is possible to see that the ansatz should have
the form
\bea \label{HVM1}
\psi^0(\vec r, t) &=& \fr{i}{\sqrt{4 \pi} }\psi_0(r,t) \ \chi \ ,  \\
\psi^i  (\vec r, t) &=& \fr{1}{\sqrt{4 \pi}} \l[ \psi_1(r,t) \ \hat r^i  + 
i \ r \ \psi_2(r,t) \ 
\left( \hat n \times \nabla^i \hat n \right) \cdot \vec \tau \r]
 \chi \ .
\label{HVM2} 
\eea 

Replacing  Eqs.(\ref{PS}), (\ref{HPS}), (\ref{HVM1}) and (\ref{HVM2}) in the effective
lagrangian, Eq.(\ref{fulllag}), we obtain that the heavy meson-soliton lagrangian $L_{HM-sol}$ is
\begin{eqnarray}
L_{HM-sol} &=& 
\frac{1}{2} \int_0^\infty \ dr \ r^2 \l\{ \dot \phi^\dagger \dot \phi -  
\phi'^\dagger \phi' - \l( m_\phi^2 
+ 2 B \frac{\cs^4}{r^2} \r) \phi^\dagger \phi 
+ \psi_0'^\dagger \psi_0' + 2 i \dot \psi_0'^\dagger \psi_1 + 
\dot \psi_1^\dagger \dot \psi_1 
\r.
\nonumber \\
& &  \l.
+ 2 B \l[ \frac{\ss^4}{r^2} \psi_0^\dagger \psi_0 +  \dot \psi_2^\dagger \dot \psi_2     
- \l(\frac{\psi_2}{r} +  \psi_2' - \frac{\ss^2}{r} \psi_1\r)^\dagger \l( \frac{\psi_2}{r} + 
\psi_2' - \frac{\ss^2}{r} \psi_1\r) 
+ 2 i \frac{\ss^2}{r} \dot \psi_0^\dagger \psi_2  \r] 
\r. 
\nonumber \\
& & \l. 
- 4 {\cal I} \frac{\cs^4}{r^2}  \psi_2^\dagger \psi_2 + 
m_\psi^2 
\l( \psi_0^\dagger \psi_0 - \psi_1^\dagger \psi_1 - 2 B \psi_2^\dagger \psi_2 \r)
- 
f \phi^\dagger \l( F' \psi_1 - 2 B \frac{s}{r} \psi_2 \r) 
\r.
\nonumber \\
& & 
\l. 
- 2 g B \l[ \frac{s}{r} ( \psi_0'^\dagger \psi_2 - \psi_0^\dagger \psi'_2 - 
\frac{\psi_0^\dagger \psi_2}{r})  + 2 s \frac{\ss^2}{r^2} \psi_0^\dagger \psi_1 -
\frac{F'}{r} \psi_2^\dagger \psi_0 + i \dot \psi_2^\dagger 
\l( 2 \frac{s}{r} \psi_1 - F' \psi_2 \r) \r] \r\} \nonumber \\
& & + h.c.
\end{eqnarray}
where $s$ and $\ss$ have been already defined, $\cs = \cos (F/2)$ and ${\cal I}$ is the angular integral 
\begin{equation}
{\cal I}  = \fr{r^4}{16 \pi} \int {d\Omega} \ \l( \nabla^i \hat n_a \nabla^i \hat n_a \r)^2.
\end{equation}

The diagonalization of the hamiltonian obtained from $L_{HM-sol}$ leads to a set of eigenvalue equations
for the heavy meson field. They are
\bea \label{eom1}
&\phi''& + \fr{2}{r} \phi' - \l(   m_\phi^2 - \om^2 + 2 B \fr{\cs^4}{r^2} \r) \phi - \fr{1}{2} f F' \psi_1 + 
f B \frac{s}{r} \psi_2 = 0 \ , \nn \\
& & \\
&\psi_0''& + \fr{2}{r} \psi_0' - \l(m_\psi^2 + 2 B  \fr{\ss^4}{r^2} \r) \psi_0
+ \l( \fr{2 \om}{r} + 2 g B   \fr{s \ss^2}{r^2} \r) \psi_1 + \om \psi_1'  
\nn \\
& & \qquad  \qquad \qquad  \qquad \qquad  \qquad
- \fr{2 B }{r} \l(g F' \cs^2 + \om \ss^2 + g \fr{s}{r} \r) \psi_2  - 2 g B  \fr{s}{r} \psi_2' = 0 \ , \nn \\
& & \\
&2 g B& \fr{s \ss^2}{r^2} \psi_0 - \om \psi_0' + \l(m^2_\psi - \om^2 + 2 B \fr{\ss^4}{r^2} \r)\psi_1 
+ \fr{1}{2} f F' \phi +  \fr{2 B}{r} \l( g \om s - \fr{\ss^2}{r} \r) \psi_2 - 2 B \fr{ \ss^2}{r} \psi_2' = 0 \ , \nn \\
& &  \\  
&\psi_2''& + \fr{2}{r} \psi_2' + f \fr{s}{2 r} \phi + \fr{\ss^2}{r} \l( g F'
+ \om \r) \psi_0 - g \fr{s}{r} \psi_0' - \fr{s}{2 r} \l(F' + 2 g \om \r)
\psi_1 - \fr{\ss^2}{r} \psi_1' \nn \\
& & \qquad  \qquad \qquad  \qquad \qquad  \qquad \qquad  \qquad
- \l( m_\psi^2 - \om^2 - g \om F' + \fr{2 {\cal I}}{B} \fr{\cs^4}{r^2} \r) \psi_2 =0 \ . \label{eom4}
\eea  

The numerical solution of this set of coupled equations supplemented with the appropiate boundary conditions
provides the heavy meson energy $\omega$ for the different baryon numbers $B$. The corresponding results
are discussed in the following section.

\section{Numerical results}

In our numerical calculations we use two sets of parameters in the $SU(2)$ sector.
Set A corresponds to the case
of massless pions and Set B to the case where the pion mass takes its
empirical value $m_\pi = 138 \ \rm{MeV}$. In both cases, $f_\pi$ and $\eps$
are adjusted so as to reproduce the empirical nucleon and $\Delta$ masses.
The fitted values are $f_\pi = 64.5 \ \rm{MeV}$ ,  $\eps = 5.45 $ for Set A and 
$f_\pi = 54 \ \rm{MeV}$ ,  $\eps = 4.84 $ for Set B. 
 With these parameters fixed  and using for 
each baryon number $B$ the rational map given in Ref.\cite{HMS98},  we obtain the profile $F(r)$  that minimizes the mass of the soliton.

We proceed to solve the bound state eigenvalue equations (\ref{eom1})-(\ref{eom4}), 
using the  values shown in Table I  for the  parameters that appear 
in the heavy meson  lagrangian, Eq.(\ref{hlag}).  
For the pseudoscalar and vector meson masses we use the empirical values. 
On the other hand, since little is known about the heavy meson coupling constants, 
we use the heavy quark symmetry relation \cite{Yan92}
\bea
f &=& 2 \ m_\psi \ g 
\eea 
as a guideline in order to estimate $f$. We take $g$ to be the value given by the non-relativistic quark model:  $g=-0.75$.
As discussed in  a  recent analysis \cite{Che97}, this value is compatible with the upper bound $g^2 \lesssim .5$ established by the experimental upper limit for  
the decay width $\Gamma(D^{*+}) < 131 \ {\rm keV} $ set by the ACCMOR Collaboration  \cite{Ba92}. 
  
Our results for the heavy meson binding energies $\veps_B$ defined by
\begin{equation}
\veps_B = m_\phi - \om_B
\end{equation}
are shown in Table II. Also listed in Table II are the soliton masses per baryon number
$M_{sol}$  taken from Ref.\cite{SS98}.

The general structure shown by the binding energies is in qualitative 
agreement with what was  found in Ref.\cite{SS98} for the strangeness 
case. For the heavy flavors we also find  that the binding energy decreases 
with increasing baryon number, except for the crossings 
that take place at $B=4$ and $B=7$. This general behaviour is the 
opposite to what was found in Ref.\cite{KoZa99}. It should 
be pointed out that, due to the absence of explicit vector mesons in the
corresponding effective action, the approach used in \cite{KoZa99} is expected to 
be less accurate than the one followed in the present work.

From Table II we also notice that, for heavy flavors, the 
binding is stronger for Set A than for Set B. This can be understood as 
follows. In the $m_Q \rightarrow \infty $ limit the heavy flavored meson 
would be concentrated at the origin \cite{JMW93}, wrapped by the 
soliton. Thus, it would only probe the potential at this point. For $B=1$ 
such potential is basically proportional to $ | g F'(0) | $ and attractive. 
As well known (see, e.g., Fig.1 of  Ref.\cite{AdNa84}), $|F'(0)|$ is larger 
in the massless case. That leads to the observed behaviour. Similar analysis
can be done for higher values of $B$. For the strangeness case the behaviour
of the profile function at medium distances becomes important and the situation
is reversed \cite{SS98}. 

In order to study the stability of the heavy multiskyrmions we
will only consider the mass of the background multiskyrmion and 
the binding energy of the bound mesons. This should be a good 
approximation since the non-adiabatic corrections are expected to be small
as a consequence of the rather large values of the moments of inertia
involved\cite{Irw98}. In the following we will focus on those states which 
have flavor number equal to their baryon number. These states are of particular
interest since in the strange sector the analog states, namely those with $Y=0$, 
have been found to be stable for some values of $B$\cite{IKS88,SS98}.

Using
\bea
I_B = M_1 + M_{B-1} - M_B
\eea
with $M_B = B (M_{sol} + m_\phi - \veps_B)$ and the values given in Table II,  
we get the ionization energies $I_B$ shown in Table III. We observe that the  only heavy flavored states that
may be stable are those with $B=4$ and $B=7$. In Table IV we summarize the energies 
for the other possible strong decays of these states. We observe that, although these states are stable against 
strong decays into two fragments, some decays into a larger number of fragments are allowed. However,
since the usual phase space factors tend to suppress the decay rates as the number of fragments $ {\it n} $
in the final state increases we expect them to be quite narrow. 
For instance, in the case of the charmed heptalambda the decay width will be very 
small since the only allowed decay mode is the one that has seven $\Lambda_c$ in the final state (Set A). 

Therefore the present model predicts, both for charm and bottom, narrow heavy multibaryon states 
with baryon number four and seven. The main reason for the unstability of
these particles can be traced back to the rather large difference that  
exists between the $B=1$ and $B>1$ binding energies. This can be easily understood
noting that, to leading order in $1/m_Q$, the only non-vanishing binding energy would be 
that corresponding to $B=1$ since for $B>1$ the radial derivative of the
soliton profile function vanishes at the origin. On the other hand,  in the case 
of strangeness the meson wavefunction is wider and, therefore, much less sensitive 
to the value of the potential at the origin. Consequently, the gap between the
binding energies of $B=1$ and $B > 1$ is much smaller and the corresponding 
multibaryon states with baryon number four and seven turn out to be absolutely stable against strong decays\cite{SS98}.

\section{Conclusions}

In this work we have studied the masses of heavy multibaryon configurations 
using the bound state approach to the  Skyrme model. This is a natural extension 
of previous work done in the strange sector. In order to consider 
the heavier  flavors of charm and bottom in a consistent way, however, it is important 
to take into account the heavy quark symmetry. 
This is accomplished by the lagrangian given in Ref. \cite{Yan92}.
An important feature of this effective  heavy meson lagrangian is that it 
contains the degrees of freedom of the heavy  scalar meson and the heavy vector 
meson explicitly, which leads to a system of four coupled equations for the 
bound state problem instead of just one equation as in Ref. \cite{SS98}.
The baryon number is carried by the soliton configuration of the light background 
fields, for which we have used the expressions in terms of 
the rational maps given in Ref.\cite{HMS98}. 

We obtained solutions for the bound state equations for $B \leq 9$. 
As in the strangeness case we find that the binding energy 
decreases with the baryon number and that the  $B=4$ and $B=7$ states are 
the most stable against strong decays. However, for the charm and bottom 
flavors these states are not absolutely stable. Still, they are expected to be 
quite narrow since only decays into final states with three or more fragments are 
energetically allowed.  We do not expect that the collective 
quantization of the soliton-meson bound system will change this picture. 
It would be important, however, to estimate the zero point energy\cite{MK91} of these 
multiskyrmion configurations. As discussed in Ref.\cite{TSW94} this
contribution may be the cause for the H particle, which appears almost at 
threshold, to be unbound. In any case, since the predicted tetralambda and heptalambda 
are more strongly bound against two particle decays than the H, this contribution is 
not expected to be so important so as to open those leading decay channels. 

\acknowledgements
The authors wish to thank the warm hospitality of the organizers and staff 
members of the INT-98-3 program of the Institute of Nuclear Theory at the 
University of Washington where this work was initiated. NNS is supported 
in part by a grant from Fundaci\'on Antorchas, Argentina,  
the grant PICT 03-00000-00133 from ANPCYT, Argentina, and CONICET, Argentina.
C.L.S. thanks the CNPq-Brasil for financial support 
through a CLAF fellowship.

\pagebreak

\begin{center}
\begin{table}
\caption{Masses   and coupling constants  for the heavy meson  lagrangian.}
\vspace*{.5cm}
\begin{tabular}{lcc}
 & \ charm \ & \ bottom \ \\
\hline
 $ m_\phi $ &  1867 MeV &  5279 MeV \\
 $ m_\psi $ &  2010 MeV &  5325 MeV \\ 
 f     & -3016 MeV  & -7988 MeV \\
 g     & -0.75 \ \ \ \ \ \ \  & -0.75  \ \ \ \ \ \ \  \\
\end{tabular}
\end{table}
\end{center}
\begin{center}
\begin{table}
\caption{Meson binding energies $\veps_B$ (in MeV) for the case of massless pions 
(Set A) and massive pions (Set B) as a function of the baryon number B. Also listed are the soliton 
masses per baryon unit $M_{sol}$ (in MeV) 
which  are taken from 
Ref. [9]. } 
\vspace{.5cm}
\begin{tabular}{lcccccc}
                   & \multicolumn{2}{c}{$M_{sol}$}
                   & \multicolumn{2}{c}{$\veps_B$(charm)  }
                   & \multicolumn{2}{c}{$\veps_B$(bottom) } \\
\cline{2-3} \cline{4-5} \cline{6-7} 
\ \   B \ \      & \ Set A \    & \  Set B \ 
                 & \ Set A \    & \  Set B \ 
                 & \ Set A \    & \  Set B \  \\ \hline
 \ \   1  \ \  &   \ \ 863 & 864         
                                           &  383  &  328 
                                           &  554  &  474    \\
 \ \   2 \ \  &   \ \ 847 & 848          
                                           &  321  &  272 
                                           &  438  &  374     \\
  \ \  3 \ \  &  \ \ 830 & 832           
                                           &  300  &  255
                                           &  408  &  351     \\
 \ \   4 \ \   &  \ \ 797 & 798          
                                           &  301  &  256
                                           &  407  &  350     \\
 \ \   5 \ \   &  \ \ 804 & 808           
                                           &  287  &  245
                                           &  391  &  339     \\
 \ \   6 \ \  &  \ \ 797 & 802          
                                           &  283  &  243
                                           &  386  &  336     \\
 \ \  7  \ \  &  \ \ 776 & 780         
                                           &  288  &  247
                                           &  392  &  341     \\ 
 \ \  8 \ \   &  \ \ 784 & 790         
                                           &  280  &  242 
                                           &  383  &  335      \\ 
 \ \  9  \ \  &  \ \ 787 & 796       
                                           &  275  &  239 
                                           &  378  &  332    \\ 
\end{tabular}
\end{table}
\end{center}
\begin{center}

\begin{table}
\caption{Ionization energies $ I_B $ (in MeV) of the  charm and bottom 
multilambdas in the case of massless pions (Set A) and massive pions (Set B).}

\vspace{.5cm}

\begin{tabular}{lcccc}
                   & \multicolumn{2}{c}{$I_B$(charm  ) }
                   & \multicolumn{2}{c}{$I_B$(bottom ) } \\
\cline{2-3}\cline{4-5}
       \ \  B  \         & \ Set A \    & \  Set B \ 
                         & \ Set A \    & \  Set B \  \\ 
\hline
  \ \ 2 &  -92 &  -80 & -200 & -168  \\
  \ \ 3 &  -58 &  -43 & -139 & -105  \\
  \ \ 4 &   86 &   99 &   15 &   41  \\
  \ \ 5 & -121 & -111 & -196 & -163  \\ 
  \ \ 6 &  -19 &   -3 &  -92 &  -61  \\
  \ \ 7 &  148 &  159 &   87 &  113  \\
  \ \ 8 & -136 & -117 & -211 & -177  \\
  \ \ 9 &  -96 &  -93 & -164 & -146  \\ 
\end{tabular}
\end{table}
\end{center}

\begin{center}
\begin{table}
\caption{Energy balance (in MeV) for the multilambda strong decays in the case of  
massless pions (Set A) and massive pions (Set B). The first column indicates 
the number of fragments {\it n }in the final state. }
\vspace*{.5cm}
\begin{tabular}{llcccc}
 & 
 & \multicolumn{2}{c}{charm  }
 & \multicolumn{2}{c}{bottom } \\
\cline{3-4}\cline{5-6}
                     {\it  n } &    & \ Set A \    & \  Set B \ 
                         & \ Set A \    & \  Set B \  \\ 
\hline
\hline
2 
&  \ $ M_{4 \Lambda} - 2 M_{2 \Lambda} $ 
&  -120 & -136 &   -76 & -104  \\
\hline
3 &  \ $ M_{4 \Lambda} - ( 2 M_{\Lambda} + M_{2 \Lambda} )  $ 
&  -28 & -56 &   124 & 64  \\
\hline
4 &  \ $ M_{4 \Lambda} -  4 M_{\Lambda}  $ 
&  64 & 24 &   324 & 232  \\
\hline
\hline
2 &  \  $ M_{7 \Lambda} - ( M_{2 \Lambda} + M_{5 \Lambda} ) $ 
& -221 & -236 &  -195 & -220 \\
&  \  $ M_{7 \Lambda} - ( M_{3 \Lambda} + M_{4 \Lambda} ) $ 
& -158 & -168 &  -138 & -162 \\
\hline
3 &  \  $ M_{7 \Lambda} - ( M_{ \Lambda} + M_{2 \Lambda} + M_{4 \Lambda} ) $ 
& -100 & -125 &  1 & -57 \\
&  \  $ M_{7 \Lambda} - ( M_{ \Lambda} + 2 M_{3 \Lambda} ) $ 
& -244 & -267 &  -153 & -203 \\
& \  $ M_{7 \Lambda} - ( 2  M_{ \Lambda} + M_{5 \Lambda} ) $ 
& -129 & -156 &  5 & -52 \\
&  \  $ M_{7 \Lambda} - ( 2  M_{ 2 \Lambda} + M_{3 \Lambda} ) $ 
& -278 & -304 &  -214 & -266 \\
\hline
4 &  \  $ M_{7 \Lambda} - (   M_{  \Lambda} + 3 M_{2 \Lambda} ) $ 
& -220 & -261 &  -75 & -161 \\
&  \  $ M_{7 \Lambda} - (  2 M_{  \Lambda} +  M_{2 \Lambda} + M_{3 \Lambda} ) $ 
& -186 & -224 &  -14 & -98 \\
&  \  $ M_{7 \Lambda} - ( 3 M_{  \Lambda} +  M_{4 \Lambda} ) $ 
& -8 & -45 &  201 & 111 \\
\hline
5 &  \  $ M_{7 \Lambda} - ( 3  M_{  \Lambda} + 2 M_{2 \Lambda} ) $ 
& -128 & -181 &  125 & 7 \\
&  \  $ M_{7 \Lambda} - ( 4  M_{  \Lambda} +  M_{ 3 \Lambda} ) $ 
& -94 & -144 &  186 & 70 \\
\hline
6 &  \  $ M_{7 \Lambda} - ( 5  M_{  \Lambda} +  M_{ 2 \Lambda} ) $ 
& -36 & -101 &  325 & 175 \\
\hline
7 & \  $ M_{7 \Lambda} - ( 7  M_{  \Lambda} ) $ 
& 56 & -21 &  525 & 343 \\
\end{tabular}
\end{table}
\end{center}


\begin{thebibliography}{99}

\bibitem{Jaf77}
R.L. Jaffe,
Phys. Rev. Lett. {\bf 38}, 195 (1977).

\bibitem{Wit84}
E. Witten, Phys. Rev. D {\bf 30}, 272 (1984).

\bibitem{GSB98}
C. Greiner and J. Schaffner-Bielich, 
in {\it Heavy Elements and Related New Phenomena}, ed. R.K. Gupta and W. Greiner, 
(World Scientific, Singapore),
{\it  nucl-th/9801062}.

\bibitem{E864}
K. Borer {\it et al.} ,
Phys. Rev. Lett. {\bf 72}, 1415 (1994);
T.A. Armstrong et al. (E864 experiment),
Phys. Rev. Lett. {\bf 79}, 3612 (1997).

\bibitem{SV98}
J. Schaffner-Bielich and A.P. Vischer,
Phys. Rev. D {\bf 57}, 4142 (1998).

\bibitem{AMS78}
A.Th.M. Aerts, P.J.G. Mulders and J.J. de Swart,
Phys. Rev. D {\bf 17}, 260 (1978);
E.P. Gilson and R.L. Jaffe,
Phys. Rev. Lett. {\bf 71}, 332 (1993).

\bibitem{CK85}
C.G. Callan and I. Klebanov,
Nucl. Phys. {\bf B262}, 365 (1985);
N.N. Scoccola, H. Nadeau, M. Nowak and M. Rho,
Phys. Lett. {\bf B201}, 425 (1988);
C.G. Callan, K. Hornbostel and I. Klebanov,
Phys. Lett. {\bf B202}, 269 (1988);
U. Blom, K. Dannbom and D.O. Riska,
Nucl. Phys. {\bf A493}, 384 (1989).

\bibitem{RRS90}
M. Rho, D.O. Riska and N.N. Scoccola,
Phys. Lett. {\bf B251}, 597 (1990);
Z. Phys. {\bf A341}, 343 (1992);
D.O. Riska and N.N. Scoccola,
Phys. Lett. {\bf B265}, 188 (1991).

\bibitem{SS98}
M. Schvellinger and N.N. Scoccola, 
Phys. Lett. {\bf B430}, 32 (1998).

\bibitem{IW88}
N. Isgur and M.B. Wise,
Phys. Rev. Lett. {\bf 66}, 1130 (1991).

\bibitem{BD92}
G. Burdman and J.F. Donoghue,
Phys. Lett. {\bf B280}, 287 (1992);
M.B. Wise, 
Phys. Rev. D {\bf 45}, 2188 (1992).

\bibitem{Yan92}
T.-M. Yan, H.-Y. Cheng, C.-Y.Cheung, G.-L. Lin, Y.C. Lin and H.-L. Yu,
Phys. Rev. D {\bf 46}, 1148 (1992). 

\bibitem{BS97}
R.A. Battye and P.M. Sutcliffe,
Phys. Rev. Lett. {\bf 79}, 363 (1997).

\bibitem{HMS98}
C.J. Houghton, N.S. Manton and P.M. Sutcliffe, 
Nucl. Phys. {\bf B510}, 507 (1998). 

\bibitem{JMW93}
E. Jenkins, A.V. Manohar and M.B. Wise,
Nucl. Phys. {\bf B396}, 27 (1993);
E. Jenkins and A.V. Manohar,
Phys. Lett. {\bf B294}, 273 (1992);
Z. Guralnik, M. Luke and A.V. Manohar,
Nucl. Phys. {\bf B390}, 474 (1993). 

\bibitem{OPM94}
Y. Oh, B.-Y. Park and D.-P. Min, 
Phys. Rev. D {\bf 49}, 4649 (1994).

\bibitem{KoZa99}
V.B. Kopeliovich and W.J. Zakrzewski, 
{\em hep-ph/9904386}.

\bibitem{Che97}
H-Y. Cheng,
Phys. Lett. {\bf B399}, 281 (1997).

\bibitem{Ba92}
S. Barlag {\it et al.},
Phys. Lett. {\bf B278}, 480 (1992). 

\bibitem{AdNa84}
G.S. Adkins and C.R. Nappi, 
Nucl. Phys. {\bf B233}, 109 (1984).

\bibitem{Irw98}
P. Irwin,
``Zero mode quantization of multi-Skyrmions'',
{\it hep-th/9804142};
 J.P. Garrahan, M. Schvellinger and N.N. Scoccola,
``Multibaryons as Symmetric Multiskyrmions'',
{\it hep-th/9906432}.

\bibitem{IKS88}
A.I. Issinskii, V.B. Kopeliovich and B.E. Shtern,
Sov. J. Nucl. Phys. {\bf 48}, 133 (1988).

\bibitem{MK91}
B. Moussallam and D. Kalafatis,
Phys. Lett. {\bf B272}, 196 (1991);
B. Moussallam, 
Ann. Phys. {\bf 225}, 264 (1993); 
F. Meier and H. Walliser, 
Phys. Rep. {\bf 289}, 383 (1997);
N.N. Scoccola and H. Walliser,
Phys. Rev. D {\bf 58}, 094037 (1998). 

\bibitem{TSW94}
G.L. Thomas, N.N. Scoccola and A. Wirzba, 
Nucl. Phys. {\bf A575}, 623 (1994).  

 
\end{thebibliography}
\end{document}